# Direct detection of pure AC spin-current by x-ray pump-probe measurements

J. Li,[1,2] L. R. Shelford,[3] P. Shafer,[4] A. Tan,[2] J. X. Deng,[2] P. S. Keatley,[3] C. Hwang,[5] E. Arenholz,[4] G. van der Laan,[6] R. J. Hicken,[3] and Z. Q. Qiu[2]

[1] *International Center for Quantum Materials, School of Physics, Peking University, Beijing 100871, China*
[2] *Dept. of Physics, Univ. of California at Berkeley, Berkeley, CA 94720, USA*
[3] *Department of Physics and Astronomy, University of Exeter, Stocker Road, Exeter, Devon, EX4 4QL, United Kingdom*
[4] *Advanced Light Source, Lawrence Berkeley National Laboratory, Berkeley, CA 94720, USA*
[5] *Korea Research Institute of Standards and Science, Yuseong, Daejeon 305-340, Republic of Korea*
[6] *Magnetic Spectroscopy Group, Diamond Light Source, Didcot, Oxfordshire, OX11 0DE, United Kingdom*

Despite recent progress in spin-current research, the detection of spin current has mostly remained indirect. By synchronizing a microwave waveform with synchrotron x-ray pulses, we use the ferromagnetic resonance of the Py ($Ni_{81}Fe_{19}$) layer in a Py/Cu/$Cu_{75}Mn_{25}$/Cu/Co multilayer to pump a pure AC spin current into the $Cu_{75}Mn_{25}$ and Co layers, and then directly probe the spin current within the $Cu_{75}Mn_{25}$ layer and the spin dynamics of the Co layer by x-ray magnetic circular dichroism. This element-resolved pump-probe measurement unambiguously identifies the AC spin current in the $Cu_{75}Mn_{25}$ layer.

The concept of spin current is of central importance in spintronics research,[1,2] having grown from the realization that a spin polarized electrical current carries not only electron charge but also electron spin that can exert a spin-transfer torque.[3,4,5] In comparison to the rapid progress made in generating spin currents by various methods,[6,7,8] their detection has remained mostly indirect, being achieved through measurement of spin-torque driven magnetization precession,[9,10] spin-current induced second-harmonic optical effects,[11] and inverse spin Hall effect (ISHE),[12,13,14] etc. Such indirect measurements may be influenced by induced magnetic order in the nonmagnetic layer at the interface which could result in ambiguous or even contradictory interpretations.[15,16,17,18,19,20,21,22] Attempts to directly measure a DC spin current by monitoring the spin polarization in a nonmagnetic material were not successful[23] until very recently when a tiny polarization of the Cu spin ($3\times10^{-5}$ $\mu_B$) was reported in a Co/Cu sample as a spin polarized electric current was injected from the Co layer into the Cu layer.[24] However, the interpretation of this result requires a careful analysis to take into account the direct polarization of the Cu by the Co at the interface. Instead of focusing on the DC component pumped by a spin-polarized electric current, it was recently proposed that a spin current pumped by the coherent precession of a ferromagnet [e.g., ferromagnetic resonance (FMR)] carries not only a time-averaged DC component but also a much larger AC component.[25] Although FMR studies have successfully demonstrated the creation of a pure spin current by spin precession in ferromagnetic (FM)/non-magnetic (NM) multilayers[10,26,27], the AC spin current has never been observed directly. ISHE measurements unfortunately exhibit a mixture of the AC spin current effect and an electrical inductance effect.[28,29,30]

In this Letter, we report an experimental study of a Py/Cu/$Cu_{75}Mn_{25}$/Cu/Co multilayer system. A pure AC spin current was pumped into the $Cu_{75}Mn_{25}$ and Co layers by exciting FMR of the ferromagnetic Py layer at 4 GHz. Using pump-probe measurements of the x-ray magnetic circular dichroism (XMCD), we unambiguously identified the AC spin precession of the spin current in the nonmagnetic $Cu_{75}Mn_{25}$ spacer layer. In addition, phase-resolved spin precession measurements revealed a characteristic bipolar phase behavior of the Co spins that is a fingerprint of spin-current driven spin precession.

The experiment was carried out on beamline 4.0.2 at the Advanced Light Source (ALS), Lawrence Berkeley National Laboratory. Static x-ray absorption spectroscopy (XAS) measurements at a grazing angle of 20° to the sample surface at the Ni, Mn, and Co 2*p* core level ($L_{2,3}$ absorption edges) were used to identify the magnetic states of the Py, $Cu_{75}Mn_{25}$, and Co layers in a Py(12nm)/Cu(3nm)/$Cu_{75}Mn_{25}$(2nm)/Cu(3nm)/Co(2.5nm) sample grown on a MgO(001) substrate, and are shown in Fig. 1. The non-zero XMCD signals (the percentage difference of the XAS for opposite magnetic field directions) at the Ni and Co edges clearly identify the ferromagnetic state of the Py and Co films. The absence of a detectable XMCD signal at the Mn $L_3$ edge at remanence confirms the nonmagnetic state of the $Cu_{75}Mn_{25}$ film, showing that the two Cu(3nm) layers completely eliminate any magnetic proximity effect[31] of the Py and Co layers on the $Cu_{75}Mn_{25}$ layer in our sample. Element-specific hysteresis loop measurements show that while the Py and Co layers exhibit the expected ferromagnetic hysteresis loops, the $Cu_{75}Mn_{25}$ layer exhibits a paramagnetic linear dependence of the XMCD signal on the magnetic field. In addition, the Py and Co films show a distinct difference in coercivity ($H_c$) and saturation field, indicating that the $Cu(3nm)/Cu_{75}Mn_{25}(2nm)/Cu(3nm)$ spacer layer prevents any static interlayer coupling between the Py and Co layers. The absence of static interlayer coupling between Py and Co is further supported by FMR measurement on Py/Cu/Co (see Supplemental Material[32]).

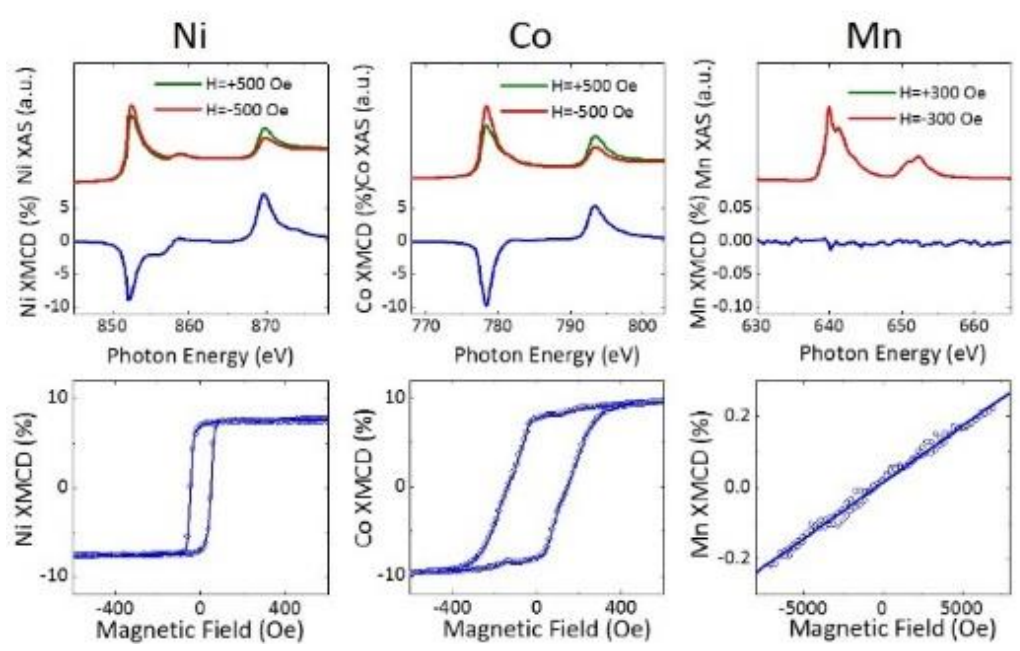

FIG. 1. (Color online) Top row: Static XMCD measurements at the Ni, Co, and Mn $L_{3,2}$ edges show that Py and Co are ferromagnetic, and the $Cu_{75}Mn_{25}$ is paramagnetic. Bottom row: Element-specific hysteresis loops obtained by monitoring field dependence of the Ni, Co and Mn $L_3$ XMCD. The Cu layers eliminate magnetic polarization and coupling of the $Cu_{75}Mn_{25}$ by the Py and Co layers.

XFMR measurements were first performed on the $Py/Cu/Cu_{75}Mn_{25}/Cu$ sample by measuring the XMCD at the Ni $L_3$ edge. By setting the time delay between the microwave RF-field (pump exciting spin precession in the sample) and the x-ray pulse (probe) to measure the absorptive (imaginary) component of the dynamic susceptibility, the pump-probe XMCD signal measures the spin precession amplitude[47,48,49]. Figure 2(a) shows the dependence of the Py spin precession amplitude as a function of applied magnetic field. The position of the Lorentzian-shaped peak shows that the Py undergoes FMR at $H_{res}$= 235 Oe for excitation at 4 GHz frequency with a full-width half-maximum linewidth equal to $\Delta H_{1/2}$=64 Oe. By changing the delay time between the microwave waveform and the x-ray pulses, the pump-probe XMCD measurement explores the full spin precession as shown by the sinusoidal shape of the XMCD signal [Fig. 2(b)]. It is clear that the spin precession exhibits a phase shift as the magnetic field is swept through the FMR resonance field.

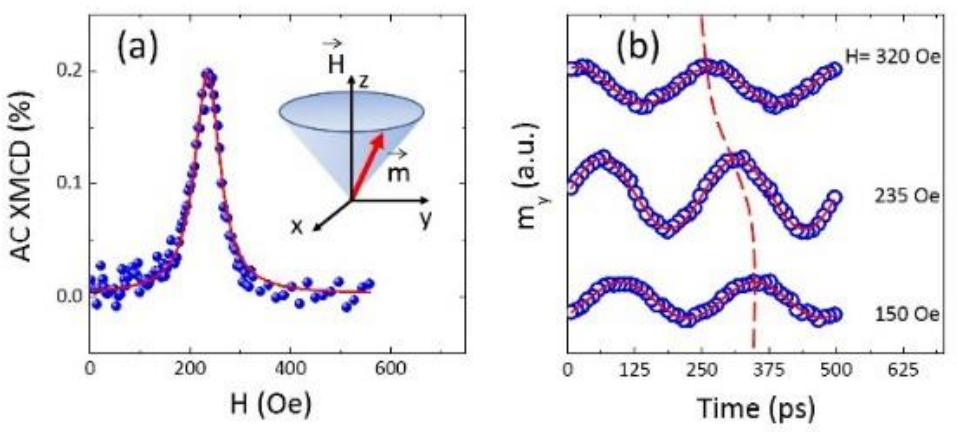

FIG. 2. (Color online) AC XMCD measurements of the Py precession in $Py/Cu/Cu_{75}Mn_{25}/Cu$. (a) The Py magnetic moment precession amplitude exhibits a Lorentzian-shaped FMR peak at $H_{res}$=235 Oe with a full-width half-maximum of $\Delta H_{1/2}$=64 Oe. (b) The sinusoidal time dependence of the Ni $L_3$ XMCD signal reveals the precession of the Py magnetic moment. A clear phase shift occurs as the magnetic field crosses the resonance field.

The spin precession of a FM layer pumps a pure spin current into a neighboring metallic layer according to

$$\vec{I}_S = \frac{\hbar}{4\pi} g^{\uparrow\downarrow} \vec{m}_{Py} \times \frac{d\vec{m}_{Py}}{dt} \, , \qquad (1)$$

where $\vec{m}_{Py} = -\vec{S}_{Py}$ is a unit vector parallel to the Py magnetic moment (antiparallel to the unit vector of Py spin $\vec{S}_{Py}$ ), and $g^{\downarrow\uparrow}$ is the dimensionless spin-mixing conductance[50]. The time-average of Eq. (1) leads to a DC spin current $\vec{I}_S^{DC} // -\langle \vec{S}_{Py} \rangle$ which is the focus of most previous works. However, a much larger AC component $\vec{I}_S^{AC} \perp \langle \vec{S}_{Py} \rangle$ can be generated by spin precession[25]. It is this spin current (unbalanced extra angular momentum) that induces a net precession spin in the direction of $\vec{I}_S$ in the nonmagnetic layer, leading to an inverted precession cone of the Cu and CuMn magnetic moments as shown in Fig. 3(a). [25,29,51] Consequently, a measurement of the Mn spin precession using XMCD at the Py FMR resonance field in our system will signify direct detection of the pure AC spin current in the nonmagnetic $Cu_{75}Mn_{25}$ spacer.

Figure 3(b) shows measurements of the Py, $Cu_{75}Mn_{25}$, and Co spin precession in the $Py/Cu/Cu_{75}Mn_{25}/Cu/Co$ sample at the Py FMR resonance field of $H_{res}$= 235 Oe for left- and right-circularly polarized x-rays. To confirm the origin of the weak Mn XMCD signal, we also performed the Mn XMCD measurement at a photon energy below the Mn $L_3$ absorption edge. The absence of any oscillations at energies below the Mn $L_3$ edge confirms that oscillatory artifacts related to RF pickup, crosstalk, and instrumental interference, etc. have been eliminated from our experiment. After careful elimination of other possible mechanisms for the Mn AC XMCD (see Supplemental Material[32]), we conclude that the observation of Mn magnetic moment precession is direct and unambiguous evidence of an AC spin current within the $Cu_{75}Mn_{25}$ layer. In particular, we present the results from the Py/MgO/CuMn sample.

From the AC and DC XMCD magnitudes, we can also estimate the magnitude of the Mn moment due to the spin current. First, we deduce the Py FMR precession cone angle from the Ni AC and static XMCD magnitudes, $\theta_{Ni} = \arctan([\text{AC XMCD (Ni)}/[\text{DC XMCD (Ni)}]) = \arctan(0.2/8) \sim 1.5°$. Then using the linear relationship between the XMCD/XAS ratio and the magnetic moment for a Mn atom,[31,52] we find that a Mn AC XMCD signal of 0.02%, as shown in Fig. 3(c), corresponds to a moment of $2.5\times10^{-3}$ $\mu_B$/Mn. The DC Mn moment due to the spin current should be ~$\tan(\theta_{Ni})\times2.5\times10^{-3}\mu_B$ = $6.5\times10^{-5}$ $\mu_B$, similar to the transient magnetic moment of $3\times10^{-5}$ $\mu_B$ reported in Ref. 24. Note this is only an estimate since the relation between magnetic moment and XMCD magnitude depends in details on the electronic structure of the material.

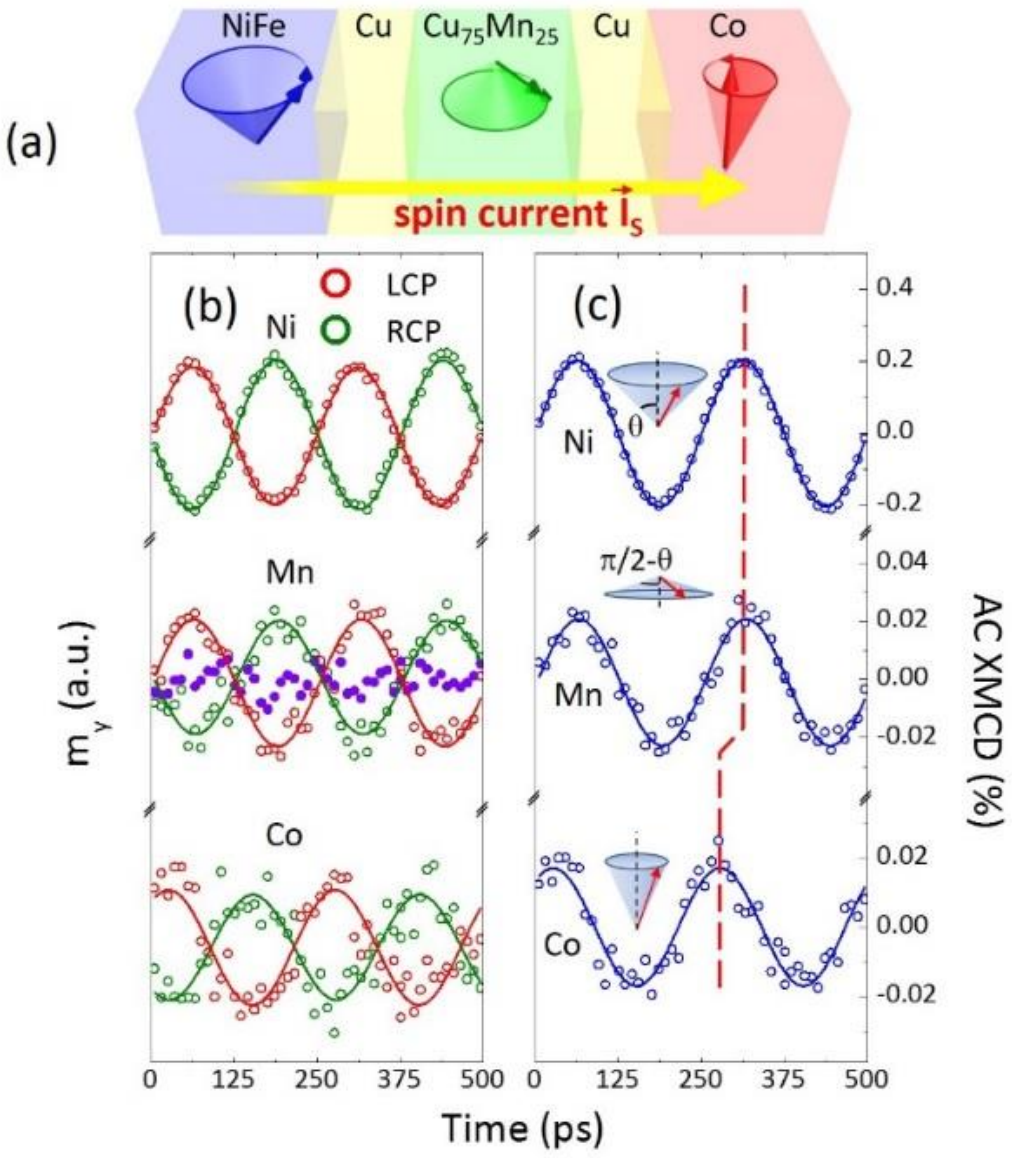

FIG. 3. (Color online) (a) Schematic drawing of the magnetic moment precession in each layer due to the pure spin current pumped by the Py FMR. Note the inverted cone of precession for the Mn moment as described by Eq. (1). (b) Spin precession within the Py, $Cu_{75}Mn_{25}$, and Co layers revealed by AC XMCD measurements using left- (LCP, red dots) and right-circularly polarized (RCP, green dots) x-rays at the Ni, Mn, and Co edges respectively. The absence of any oscillations below the Mn $L_3$ edge energy (purple solid dots) confirms the absence of any artifacts in the measurement. (c) The relative magnitude and phase of the Py, $Cu_{75}Mn_{25}$, and Co spin precession. The $Cu_{75}Mn_{25}$ spin precession is a direct indicator of the AC spin current.

We rule out electron spin resonance (ESR)[53] from the $Cu_{75}Mn_{25}$ layer. At $f$=4GHz, ESR occurs at $H\approx1300$ Oe, thus we do not expect any detectable Mn ESR signal at the Py FMR field of $H\approx230$ Oe. We proved the absence of ESR at the Py resonance field by performing time-resolved XMCD measurements on the Py(12nm)/MgO(3.0nm)/$Cu_{75}Mn_{25}$(2.0nm) sample. The insulating MgO layer blocks the spin current from the Py layer into the $Cu_{75}Mn_{25}$ layer. While the Py exhibits the expected FMR spin precession [Fig. 4(a)], no Mn AC XMCD signal is detected in the $Cu_{75}Mn_{25}$ layer at a sensitivity of 0.01% [Fig. 4(b)]. The total power absorption indicates the presence of a broad ESR peak [Fig. 4(c)] with contributions from all conducting elements in the sample (e.g., the CPW and Cu). However, no detectable Mn AC XMCD signal was found at H=1300 Oe. Therefore the Mn precession in Fig. 3 cannot be attributed to ESR or dipolar coupling between Py and Mn, but rather to the FMR of Py, which drives the Mn precession in phase with the Py (AC spin current across the Cu layer).

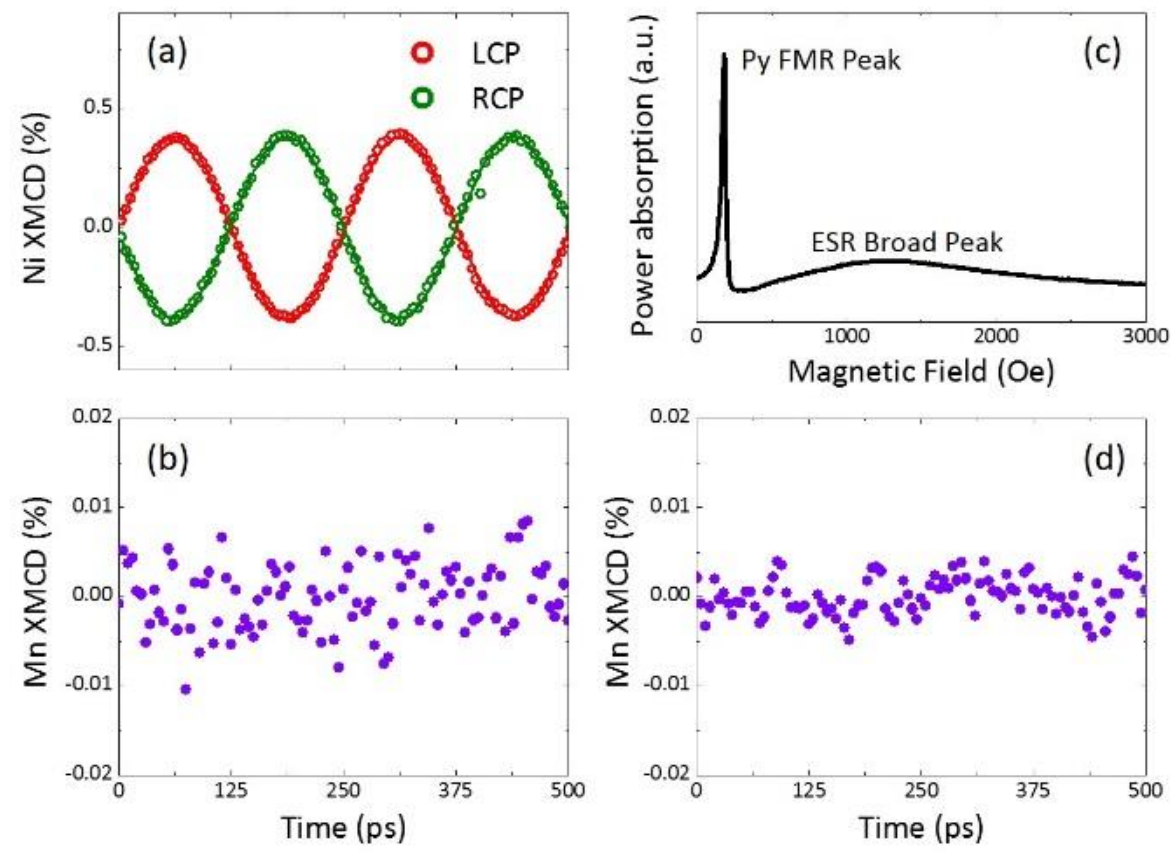

FIG. 4. (Color online) For the Py/MgO/$Cu_{75}Mn_{25}$ sample, (a) Ni spin precession at the Py resonance field. (b) Absence of Mn XMCD indicates the absence of the Mn spin precession at the Py resonance field. (c) Total power absorption showing a broad ESR peak at $H$=1300 Oe in addition to the sharp Py FMR peak. The ESR arises from all conduction electrons in the sample. (d) The absence of Mn AC XMCD at $H$=1300 Oe shows that the ESR does not contribute to the Mn AC XMCD signal.

From the pump-probe XMCD measurement, we also determined the relative phase of the Py, $Cu_{75}Mn_{25}$, and Co magnetic moment precession at the Py FMR resonance field. Figure 3(c) shows that the $Cu_{75}Mn_{25}$ magnetic moment has identical phase to the Py magnetic moment. In fact the identical phase of the Mn and Py precessions is an important property of the AC spin current in Eq. (1) (i.e., the pumped magnetic current is in phase with the pumping FMR magnetic moment).[51] In contrast, the Co magnetic moment precession has an obviously different phase to the Py magnetic moment precession. This is a clear indication that the Co magnetic moment precession cannot be explained by direct exchange coupling of the Py and Co layer through pin holes, etc. Then an interesting question is why there is a phase difference between the spin current and the Co spin precession?

We systematically measured the Py and Co precessions at different magnetic fields [Fig. 5(a)] from which the Py and Co amplitude [Fig. 5(b)] and phase [Fig. 5(c)] were extracted by fitting of the XMCD signal to a sine wave. Note the amplitudes are normalized in Fig. 5(a) for clarity. The extracted component of the Py amplitude projected onto the *y*-axis, i.e., perpendicular to the applied field, exhibits a Lorentzian-shaped FMR peak at the same resonance field of $H_{res}$=235 Oe as in Py/Cu/$Cu_{75}Mn_{25}$/Cu [Fig. 2(a)]. However, the linewidth of $\Delta H_{1/2}$=95 Oe in Py/Cu/$Cu_{75}Mn_{25}$/Cu/Co is larger than that of $\Delta H_{1/2}$=64 Oe in Py/Cu/$Cu_{75}Mn_{25}$/Cu [Fig. 2(a)], suggesting that a spin current has been pumped into the Co layer. In addition, the linewidth of $\Delta H_{1/2}$ ~ 50 Oe in Cu/Py/Cu sample at 4GHz, which is smaller than that in Py/Cu/$Cu_{75}Mn_{25}$/Cu sample, shows the existence of spin damping in the CuMn layer.

Indeed, we observe a peak in the Co magnetic moment precession amplitude right at the Py FMR field [Fig. 5(b)]. Since an isolated single Co layer has a smaller FMR resonance field, and since the spacer layer in our sample prevents any static Py-Co interlayer coupling (see Supplemental Material[32]), the Co peak at the Py FMR field must be associated with the spin current pumped by the Py FMR. Note that spin precession by a spin-polarized electrical current has previously been demonstrated in spin-torque nano-oscillators (STNOs).[9,54] Applying this idea to a $FM_1$/NM/$FM_2$ trilayer suggests that a DC spin current generated by FMR in $FM_1$ could cause the spin precession in $FM_2$. However, this scenario cannot explain our data because under these conditions the $FM_2$ spins should precess at the $FM_2$ FMR resonance field rather than at the $FM_1$ FMR resonance field. The fact that the Co peak in Fig. 5(b) appears at exactly the Py FMR field suggests that the Co peak is driven by the AC spin current rather than by the DC spin current.

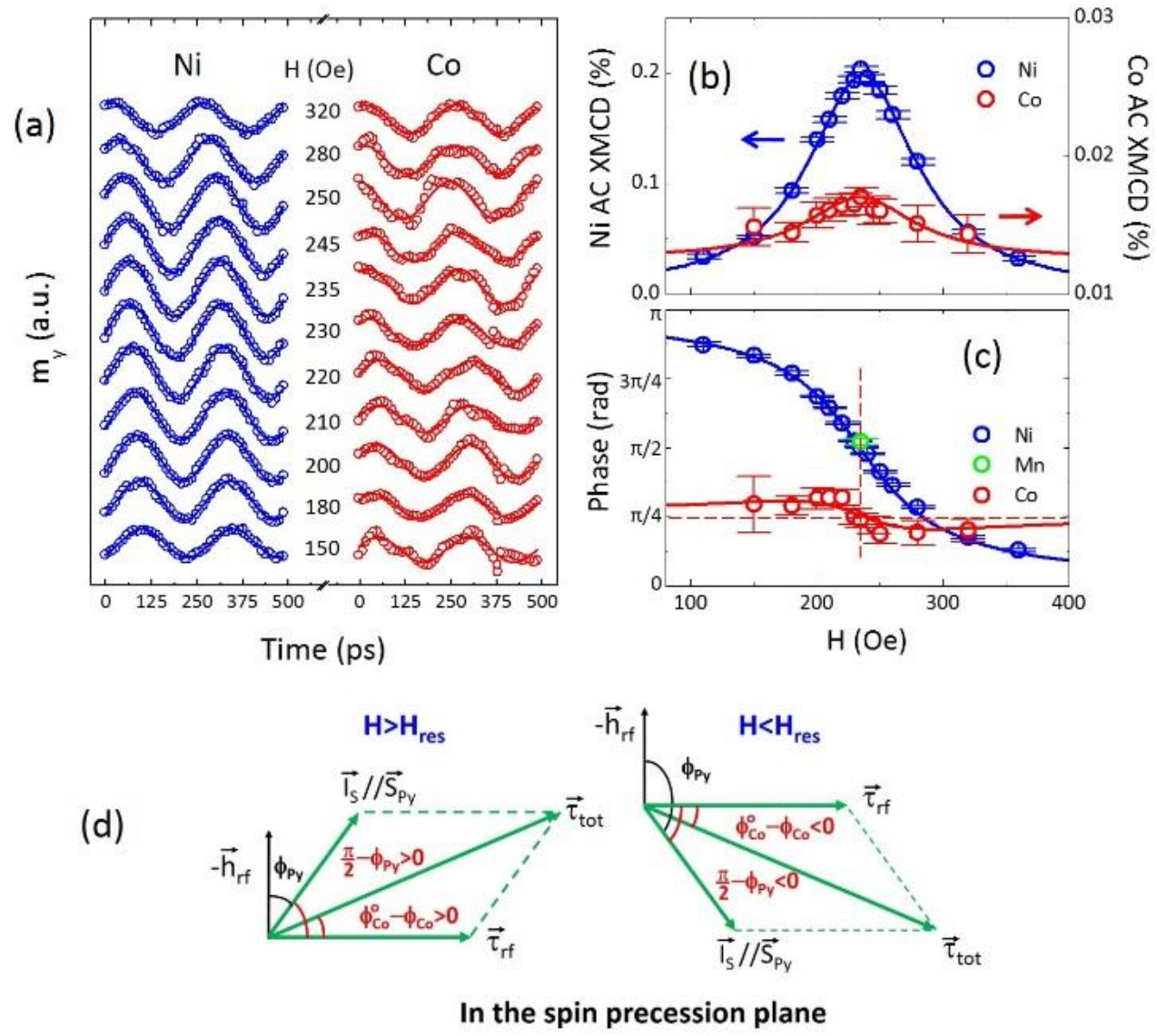

FIG. 5. (Color online) (a) Py and Co magnetic moment precession at different magnetic fields (dots are experimental data, lines are sinusoidal fits). The amplitude is normalized for clarity. (b) Ni and Co AC XMCD as a function of applied field. At the Py FMR field of $H_{res}$= 235 Oe, the Co amplitude also shows a peak due to spin pumping. (c) Phase of the AC XMCD signals. The Py precession shows the π-phase change typical of FMR across the resonance field. The phase of the $Cu_{75}Mn_{25}$ is identical to that of Py as indicated by Eq. (1). The Co phase exhibits a characteristic bipolar behavior that is a fingerprint of AC spin-current driven precession. The solid lines in (b) and (c) are calculated results (see Supplemental Material[32]). (d) From the schematic diagram of the AC spin current, RF-field torque $\vec{\tau}_{rf}$, and the total torque $\vec{\tau}_{tot}$, in the spin precession plane, it is easy to understand the bipolar phase variation, whereby $\phi_{Co} < \phi_{Co}^{0}$ for $H$>$H_{res}$ and $\phi_{Co} > \phi_{Co}^{0}$ for $H$<$H_{res}$ (see main text).

The phases of the Py and Co spin precession are shown in Fig. 5(c) together with that of Mn at the Py FMR field of $H_{res}$ = 235 Oe. The small Mn XMCD signal makes it impractical to obtain its dependence over the full field range. As the magnetic field is swept through the resonance field of $H_{res}$ = 235 Oe, the Py phase undergoes a π-phase shift typical of FMR. The Co phase, on the other hand, exhibits an obvious bipolar behavior[55] with the phase value being smaller at $H$>$H_{res}$ and larger at $H$<$H_{res}$ than for a single isolated Co layer (horizontal dotted line). This bipolar character of the Co phase variation cannot be attributed to technical issues (e.g., a constant phase offset due to the use of a doped Si substrate)[56] but on the contrary, manifests the existence of a spin torque due to AC spin current. To understand the phase behavior, recall that the phase ϕ in FMR (traditionally defined as the angle of the exciting RF-field vector relative to the magnetic moment vector in the spin precession plane) has the physical meaning that the angle π/2−ϕ is the angle between the rotating spin and the RF-field torque in the precession plane. At $H$ = $H_{res}$, the Larmor frequency of the Py is exactly equal to the microwave frequency of 4 GHz and the RF-field torque acts fully to open the FMR cone angle ($\pi/2-\phi_{Py}=0$ or $\phi_{Py}=\pi/2$). At $H$>$H_{res}$, the Py Larmor frequency is greater than 4 GHz. Therefore the RF-field torque must have a component antiparallel to the direction of precession of the Py spins ($\pi/2-\phi_{Py}>0$ or $\phi_{Py}<\pi/2$) so as to slow down the Py precession to 4 GHz [Fig. 4(d)]. Similar reasoning explains the case $\pi/2-\phi_{Py}<0$ ($\phi_{Py}>\pi/2$) at $H$<$H_{res}$. For the Co layer, the Co spin precession driven by the RF-field alone would lead to an almost field-independent phase $\phi_{Co}^{0}$ in the vicinity of the Py FMR. In the presence of the AC spin current as described by Eq. (1), the Co spin precession is driven by the total torque ($\vec{\tau}_{tot}$) due to the RF-field torque plus the AC spin current. Therefore the Co phase must take a new value $\phi_{Co}$ accounting for the change from the RF-field torque direction to the total torque direction [Fig. 5(d)]. Recall that the AC spin current has the same phase as the precessing Py spin. Then for $H$>$H_{res}$, the fact that the AC spin current vector rotates 'in advance' of the RF-field torque vector ($\pi/2-\phi_{Py}>0$) leads to a total torque that rotates 'in advance' of the RF-field torque, leading to $\phi_{Co}^{0}-\phi_{Co}>0$ or $\phi_{Co}<\phi_{Co}^{0}$ [Fig. 5(d)]. Similarly for $H<H_{res}$, the fact that the AC spin current vector lags the RF-field torque vector ($\pi/2-\phi_{Py}<0$) leads to the total torque vector lagging behind the RF-field torque direction, leading to $\phi_{Co}^{0}-\phi_{Co}<0$ or $\phi_{Co}>\phi_{Co}^{0}$ [Fig. 5(d)]. This is exactly the bipolar behavior observed in our experiment. A detailed analysis (Supplementary Material[32]) explains this bipolar behavior quantitatively [red solid line in Fig. 5(c)]. In contrast, a static Py-Co interlayer coupling torque $\sim \vec{S}_{Py}\times\vec{S}_{Co}$ causes the precessing Py spin to behave as an effective RF-field rather than as an RF-field torque, leading to only a unipolar variation of the Co precession phase.[57]

In summary, we have investigated the spin pumping effect in $Py/Cu/Cu_{75}Mn_{25}/Cu/Co$. The Py FMR pumps a pure spin current into the $Cu/Cu_{75}Mn_{25}/Cu$ spacer layer and generates precession of the Co spin. We performed pump-probe XMCD measurements to observe element-specific Py, $Cu_{75}Mn_{25}$, and Co spin precession. We directly observed the AC spin current by detecting the $Cu_{75}Mn_{25}$ spin precession. The AC spin current has the same phase as the Py spin precession and excites precession of the Co spin at the same frequency but with a different phase. The fact that the AC spin current has the same phase as the Py spin precession leads to the characteristic bipolar phase behavior of the Co spin precession. Our experiment not only directly identifies the AC spin current in the non-magnetic spacer layer, but also shows how the AC spin current transfers its angular momentum so as to generate the Co spin precession.

**Acknowledgements**

We acknowledge helpful discussion with Arne Brataas. Financial support from National Science Foundation DMR-1504568, Future Materials Discovery Program through the National Research Foundation of Korea (No. 2015M3D1A1070467), and Science Research Center Program through the National Research Foundation of Korea (No. 2015R1A5A1009962) is gratefully acknowledged. The Advanced Light Source is supported by the US Department of Energy under contract number DE-AC02-05CH11231. JD acknowledges fellowship support from the China Scholarship Council and National Science Foundation of China under Grant Nos. 51331006. LRS, PSK, and RJH acknowledge the support of the Engineering and Physical Sciences Research Council (EPSRC) through grants EP/J018767/1 and EP/I038470/1. GvdL acknowledges support of the EPSRC through grant EP/J018767/1.